# Regular "Breathing" of Single-Cycle Light Bullets in Mid-IR Filament.

S.V. Chekalin,[1,*] V.O. Kompanets,[1] A.V. Kuznetzov,[2] A.E. Dormidonov,[3,†] S.A. Shlenov,[3] and V.P. Kandidov[3]

[1]*Institute of Spectroscopy RAS, 142190, Troitsk, Moscow, Russia*
[2]*Irkutsk Branch of Institute of Laser Physics SB RAS, 664033, Irkutsk, Russia*
[3]*Moscow Lomonosov State University, Physics Department and International Laser Center, 119991, Moscow, Russia*
(20.11.2015)

Experimental and numerical studies of a temporal evolution of a light bullet formed in isotropic LiF by Mid-IR femtosecond pulse (2500 – 3250 nm) of power, slightly exceeding the critical power for self-focusing, are presented. For the first time regular oscillations of the light bullet intensity during its propagation in a filament were registered by investigation of induced color centers in LiF. It was revealed that color centers in a single laser pulse filament have a strictly periodic structure with a length of separate sections about 30 $\mu$m, which increases with a laser pulse wavelength decreasing. It was shown that the origin of light bullet modulation is a periodical change of the light field amplitude of an extremely compressed single-cycle wave packet in a filament, due to the difference of the wave packet group velocity and the carrier wave phase velocity.

DOI:                              PACS numbers:                           .

One of the most interesting of filamentation regimes is a process of a "light bullet" (LB) formation, when the simultaneous self-focusing of the beam and self-compression of a laser pulse occur. The conception of the LB formation due to the nonlinear optical self-action of a laser pulse in a dispersive medium with a cubic nonlinearity was the first formulated by Silberberg [1] from an analysis of the quasi-optical equation in the aberrationless paraxial approximation. A promising approach for the pulse self-compression is to use the anomalous group velocity dispersion (GVD) in the presence of positive self-phase modulation (SPM) that could cause the formation of an extremely compressed wave packet. In the case of anomalous GVD the light field is contracted to the central temporal slices of the pulse from its edges [2]. Pulse compression in time and space due to the Kerr nonlinearity, self-phase modulation at anomalous GVD, self-focusing, self-steepening, and defocusing of the pulse tail in laser-induced plasma creates a LB in a filament [3-5]. In the case of 1800 nm pulse filamentation in fused silica the peak intensity in a LB achieves the value of $5 \times 10^{13}$ W/cm$^2$, its diameter is about 20 $\mu$m, and the duration becomes less than two oscillation cycles of the light field for the centre wavelength, a temporal profile being considerably different from the Gaussian. The evolution of an 1800-nm LB temporal-spatial structure in sapphire was investigated in [6]. One of the questions under the discussion concerns the LB propagation behavior in a nonlinear medium: either quasistationary [7] or recurrent [3-5]. In [7] a long glowing channel of a 1900 nm pulse filament is considered as a long-lived quasi-soliton, which does not change a shape over the distance of a several centimeters. The opportunity for sequence formation of quasi-periodical LBs under filamentation in anomalous GVD regime was investigated numerically in [8]. The short-lived LBs sequence formation has been confirmed numerically and experimentally by autocorrelation measurements in [4,5,9]. LBs are robust structures and the formation of each LB in the sequence is accompanied by the outburst of certain portion of SC energy in the visible range [10,11].

An important and still poorly investigated aspect of filamentation is related to the compressed wave packet carrier envelope phase-induced effects along the few-cycle pulses propagation in a transparent medium [12,13].

In this letter we present the results of experimental and numerical studies of a temporal and spatial structure of a single LB formed in isotropic material LiF by a Mid-IR (2500 – 3250 nm) femtosecond pulse of power slightly exceeding the critical power for self-focusing. For the first time regular oscillations of LB intensity during its propagation in isotropic crystal were recorded. We used laser coloration method that gives a unique possibility to restore the filament spatial structure after exposure [14]. This experimental method, based on microscopic observation of laser-induced CCs, allowed to register a strictly periodic structure with a length of separate sections about 30 $\mu$m in a ~ 1 mm length filament created by a single laser pulse. Numerical simulation showed that the origin of the recorded CCs periodic structure is the light field amplitude oscillations of an extremely compressed single-cycle LB in a Mid-IR filament.

The pulses at wavelengths of 2500 – 3250 nm, corresponding to the region of anomalous dispersion in lithium fluoride (zero GVD wavelength is 1234 nm), were generated by a travelling-wave optical parametric amplifier of white-light continuum (TOPAS-C) with a noncollinear difference frequency generator (NDFG). The FWHM duration of transform-limited 3000 nm pulses was equal to 100 fs with spectral width 200 – 250 nm (FWHM). Pulse repetition rate was set to 100 Hz, the pulse energy could be varied from 0.2 to 16 $\mu$J. The laser pulses were focused inside an isotropic LiF sample by a thin CaF$_2$ lens with a focal length $F = 10$ cm, geometrical focus was at 3 mm from the sample input face. During a coloration process, the sample was moved in the direction perpendicular to the laser beam to insure one-shot exposure of every filament. After the end of coloration a spatial distribution of laser-induced CCs luminescence along filaments has been measured by microscope Euromex Oxion and the confocal fluorescence microscope PicoQuant MicroTime 200.

LiF has been selected in our experiments because of much more bright CC luminescence intensity in comparison with other alkali halide crystals that allowed us to observe a photoinduced transformation produced by a single laser pulse. The mechanisms which led to coloration may be explained by the non-linear excitation of electrons to the conduction band via different processes such as avalanche ionization, tunnelling ionization, and multiphoton absorption [15,16] and also by a direct multiphoton excitation of excitonic band, which is possible only for Mid-IR exciting pulses.

The photos of CCs luminescence in filaments produced by 3000-nm laser pulses are shown in Fig. 1. CC luminescence excited by 450 nm CW laser was recorded by a digital camera Nikon D800. It can be seen that the single pulse filaments have a strictly periodic structure with a length of separate sections about 30 $\mu$m. The characteristic length of the CC structures in the filaments is noticeably less than 1 mm and their thickness is about several microns i.e. of the order of the exciting laser wavelength. These parameters prove LB formation because it is a typical size of a LB [11]. The filament start distance is slightly changed from shot to shot due to laser energy fluctuations. That is why all structure is randomly moved along the direction of the laser pulse propagation.

One more characteristic feature of the observed CC microstructures is their dependence on the excitation wavelength (Fig. 2). A processing of more than 25 CC structures profiles in every set of experiments with different excitation wavelengths resulted in a conclusion that the structure period is constant at given wavelength. The CC structure induced in filament does not depend on the laser pulse intensity — its period is the same in filaments, observed at the exciting laser energies changing from 8.5 $\mu$J (threshold of CCs formation) to 16.5 $\mu$J. It also does not depend on focusing conditions (it is the same for focusing by 10 cm and 15 cm lens) and on the orientation of the sample with reference to the direction of exciting laser beam. The recorded structures were scanned in the microscope with the help of a piezo-scanner translated with an accuracy of 0.8 $\mu$m.

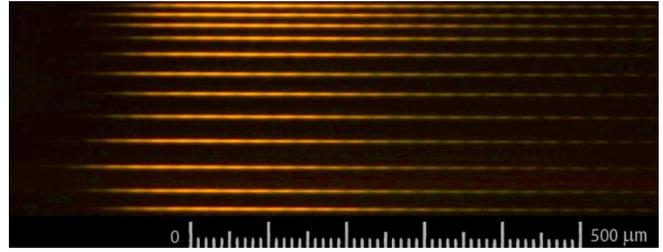

FIG. 1. Photoimages of CC structures induced in LiF under filamentation of 3000 nm laser pulse with the energy of 15.5 $\mu$J. The laser beam direction – from the left to the right. The longitudinal scale (division value is 10 $\mu$m) denotes the distance from the sample input face.

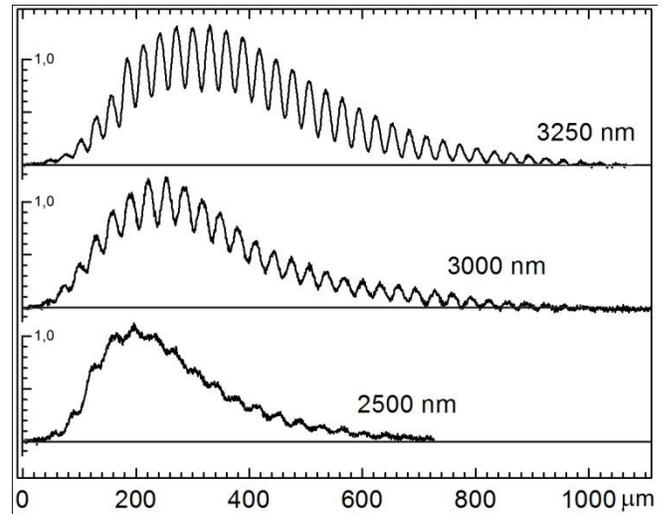

FIG. 2. Luminescence intensity of CC structures (in arbitrary units) induced in LiF under filamentation of a single 100 fs laser pulse at different wavelengths at various distances along the filament. The laser pulse parameters: wavelength 2500 nm, pulse energy 10.1 $\mu$J; wavelength 3000 nm, pulse energy 11.5 $\mu$J; wavelength 3250 nm, pulse energy 13.5 $\mu$J.

One can see that a luminescence signal modulation decreases and a period increases with the excitation wavelength decrease. The CC microstructures period is equal to 36.5 ± 0.5 $\mu$m, 31.1 ± 0.1 $\mu$m, and 29.0 ± 0.05 $\mu$m for the wavelengths of 2500, 3000, and 3250 nm, correspondingly. A modulation was not observed at wavelengths less than 2500 nm. From the other hand, we cannot observe filaments at wavelengths more than 3250 nm due to deficient femtosecond laser pulse power and quadratic increase of critical power for self-focusing with wavelength.

Average measured values of diameter of the structures induced by the laser pulses at 2500, 3000, and 3250 nm are about 1.9, 3.0, and 3.2 $\mu$m respectively. Thus, the values are comparable to the wavelengths of the excitation pulses.

Numerical simulations clearly illustrate the scenario of the formation of LBs temporal structure, which results in appearing of the observed CC density modulation. Our simulations were performed in the slowly varying wave approximation [17]. In the formulation of the problem under consideration [11], we take into account the diffraction and dispersion of a wave packet, the Kerr self-focusing, photo- and avalanche ionization of the medium, the aberrational defocusing and absorption of the wave packet in the laser-induced plasma, as well as the effect of front self-steepening of the pulse envelope. In our numerical model, the LiF dispersion was determined by the Sellmeier equation and the photoionization rate by the Keldysh formula [18].

The initial laser pulse parameters in simulations are close to the experimental values: pulse duration — 100 fs (FWHM), energy — 10.5, 15.5, and 18 $\mu J$ for pulses at 2500, 3000, and 3250 nm, respectively, that corresponds to the pulse peak power about 1.5 $P_{cr}$, where $P_{cr}$ — the critical power for self-focusing.

Calculated light field intensity distributions $I(r,\tau) \sim |A(r,\tau)|^2$ of 3000-nm pulse at distances 7.26 and 7.59 mm after the entrance in the propagation medium are shown in Fig. 3 in the pulse local frame $\tau = t - z/v_g$ where $\tau$ denotes the retarded time, $v_g$ — pulse group velocity. We use a logarithmic color scale for expanding a range of reproducible intensity levels up to four orders of magnitude. At the distance $z = 7.26$ mm [Fig. 3(a)] the pulse peak intensity starts to grow rapidly — the LB formation begins. At the distance $z = 7.56$ mm the pulse peak intensity reaches the value $I \approx 2 \times 10^{14}$ W/cm$^2$ (150 $I_0$, the initial intensity) that is enough for the multiphoton and avalanche ionization and excitons generation [19] in the medium. In the strong anomalous GVD regime the light field intensity increases due to both spatial self-focusing and temporal pulse self-compression, which leads to the formation of an extreme compressed LB [10,11] [Fig. 3(b)]. The LB duration becomes about 10 fs (FWHM) that is about 10 % of the initial pulse duration. So, the formed LB is a stable wave packet comprising only one light field oscillation that propagates in the medium with a pulse group velocity at distances of several hundred microns.

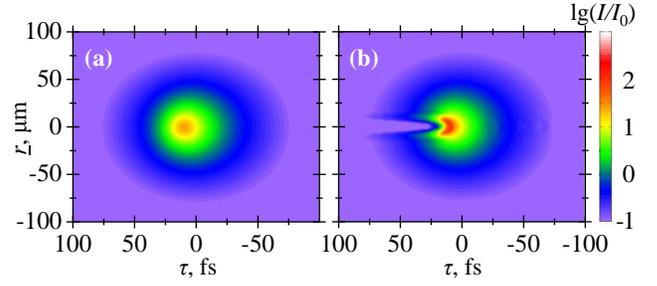

Fig. 3 (color online). Spatial-temporal intensity distribution of LB formed by 100-fs pulse at 3000 nm central wavelength with an energy of 15.5 $\mu J$. (a) The beginning of LB formation is at $z = 7.26$ mm, its peak intensity is $3 \times 10^{14}$ W/cm$^2$, duration — 25 fs (FWHM), (b) LB is formed at $z = 7.56$ mm, its peak intensity is $2 \times 10^{14}$ W/cm$^2$, duration 10 fs (FWHM).

The light field $E(r,z,\tau)$ in LB wave packet can be represented as [17,20]

$$E(r,z,\tau) = \tfrac{1}{2} A(r,z,\tau) \exp\left\{i\omega_0\tau + iz\omega_0\left(\tfrac{1}{v_g} - \tfrac{1}{v_{ph}}\right)\right\} + c.c., \quad (1)$$

where $A(r,z,\tau)$ — calculated pulse envelope, $\omega_0 = 2\pi c/\lambda_0$ — pulse carrier frequency, $v_{ph} = c_0/n(\lambda_0)$ — phase velocity. This equation describes the evolution of the light field wave packet under the change of envelope $A(r,z,\tau)$ and phase $\varphi(z) = z\omega_0\left(\tfrac{1}{v_g} - \tfrac{1}{v_{ph}}\right)$, which determines the temporal shift of the carrier oscillations with respect to the envelope peak. The complex envelope $A(r,z,\tau)$ evolves due to dispersion, diffraction, nonlinearities, and absorption, whereas $\varphi(z)$ evolves due to a difference between the wave packet group velocity $v_g$ and phase velocity $v_{ph}$ at the carrier frequency $\omega_0$ in the propagation medium.

Fig. 4(a) shows the light field at the wave packet axis $E(r=0,\tau)$ at the distance $z = 7.26$ mm corresponding to the initial stage of LB formation [see Fig. 3(a)]. The pulse envelope contains a few cycles of the carrier wave and the peak amplitude of the electric field increases in five times. At $z \approx 7.56$ mm the pulse extreme compression occurs and the single-cycle wave packet is forming [Figs. 4(b)–4(d)]. Due to the difference $\Delta v = v_{ph} - v_g$ the carrier wave moves faster than the wave packet envelope.

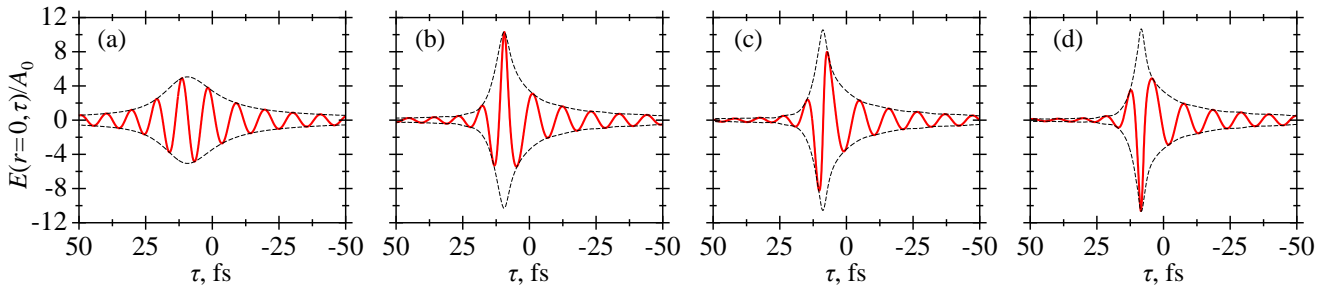

Fig. 4 (color online). Light field profiles on the axis of LB at several distances z: (a) $z = 7.26$ mm — initial stage of LB formation; (b) $z = 7.559$ mm — single-cycle LB envelop maximum matches a maximum of carrier wave; (c) $z = 7.578$ mm — single-cycle LB envelop maximum matches a zero of carrier wave; (d) $z = 7.596$ mm — single-cycle LB envelop maximum matches a minimum of carrier wave.

So, if at $z_1 = 7.559$ mm [Fig. 4(b)] a maximum of the carrier wave coincides with the maximum of the pulse envelope, at $z_2 = 7.579$ mm [Fig. 4(c)] the carrier wave shifts at quarter period and the resulting peak amplitude of the electric field in LB decreases by more than 20 %. At $z_3 = 7.596$ mm [Fig. 4(d)] the pulse envelop maximum matches a minimum of carrier wave. The electric field amplitude in the LB becomes maximal again. By this way under LB propagation as a single-cycle wave packet, the light field amplitude "breathes" due to a difference in the wave packet group velocity and the phase velocity of the carrier wave. These oscillations of electric field peak amplitude proceed throughout all the LB's existence.

For the pulse at 3000 nm the energy gap between the strongly bound singlet exciton state and the valence band (12.8 eV [21]) is equal to the energy of more than 30 Mid-IR laser quanta. That is why a relatively small change of the light field results in a considerable oscillation of CC density. This regular "breathing" of the light field in LB due to $\varphi(z)$ change during its propagation in the media results in CC periodic structure arising under Mid-IR filamentation in LiF. The phase $\varphi(z)$ impact on nonlinear-optical interaction of a single-cycle laser pulse with media at characteristic interaction time comparable to the period of optical oscillation was discussed in [20]. The period of peak amplitude oscillations in LB of 3000-nm pulse obtained from the numerical simulation is equal to $\Delta z = \overline{z_3 - z_1} = 37 \pm 0.5\,\mu m$. For the pulse at 3250 nm LB breathing period $\Delta z$ decreases to $35 \pm 0.5\,\mu m$ nm, at 2500 nm $\Delta z$ increases to $44 \pm 1\,\mu m$.

A simple analytical estimation of an oscillations period can be derived from a round-trip carrier phase shift relative to the envelope. The equation $\varphi(\Delta z^*) = \pi$, which does not account for nonlinearity action on the wave packet phase modulation and group velocity, gives the expression:

$$z_{simple} = \frac{\lambda_0}{2n(\lambda_0)\Delta v} v_g, \qquad (2)$$

For the pulse at $\lambda_0 = 3000$ nm (2) gives $\Delta z^* = 34\,\mu m$ that coincides very well with numerical simulations. For pulses at $\lambda_0 = 2500$ nm and $\lambda_0 = 3250$ nm $\Delta z^*$ equals to $41\,\mu m$ and $31.5\,\mu m$, respectively. These results perfectly match the experimentally measured period of CC structure in LiF formed in Mid-IR filament.

The periodic structures in Mid-IR filament observed in isotropic LiF have a quite another origin than those reported beforehand under an investigation of filamentation in different media. The observed behavior is opposite to a periodic filamentation of laser femtosecond pulses observed in recent experiments with birefringent crystals, which is caused by the periodic change in the polarization of the pulse travelling in birefringent medium in combination with the cross-sectional difference in multiphoton absorption for the linear and circular polarizations [22]. A period of a single filament structure in these experiments increased with the excitation pulse wavelength increasing in contrast to our measurements with isotropic crystal LiF. In the filamentation regime close to self-guiding at normal GVD investigated numerically in [23] for condensed matter and in [24] for air the period of the light field oscillations caused by manifold refocusing decreases with decreasing of the laser pulse power in respect to the critical power for self-focusing. In quasi-periodical LBs sequence formed under filamentation at anomalous GVD the interval between the bullets is not strictly periodic, its value being much more than those recorded in the present experiments. The physical reason for the CC density modulation is a cyclic transformation of the light field in a single Mid-IR LB travelling in LiF, caused by the change of the phase shift between the carrier wave and the wave packet due to the difference in a LB group velocity and a phase velocity of the carrier wave. By this way under LB propagation as a single-cycle wave packet, the light field amplitude "breathes". These relatively small oscillations of the light field peak amplitude result in a considerable oscillation of CC density due to a high nonlinearity power of multiphoton process of CC creation.

In conclusion, a regular "breathing" of the light field in LB formed under Mid-IR filamentation has been revealed, which proceeds throughout all the LB's existence. This periodic light field oscillation was discovered by investigation of periodic CC structures created in one pass of LB through an isotropic crystal LiF. For filamentation at 3000 nm the CC structure period induced in filament is about 30 μm and decreases with the excitation wavelength increasing. It also does not depend on the laser pulse power, if it is greater than the critical power for self-focusing, on focusing conditions and on the orientation of the sample. Mid-IR LB is a single-cycle wave packet and a regular change of its light field maximal amplitude in the course of propagation in filament is the result of a periodical change of the phase shift between the carrier wave and the wave packet envelope due to a difference in the LB group and phase velocities. Stability of a photoinduced periodic CC structure confirms LB robustness, which is the result of intense light field self-organization under its nonlinear-optical interaction with medium in anomalous GVD condition [11].

Tracing of periodic CC structures in LiF is a way of registration of the absolute carrier wave phase change in one-cycle Mid-IR wave packet that allows scale attosecond physics investigations on phase sensitivity of nonlinear-optical interactions. The possibility of such a recording in the IR optical range was discussed in review [12].

This work is supported by grants RFBR 14-22-02025-ofi_m, 15-32-50193-mol_nr and Grant of the President of the Russian Federation NSh-3796.2014.2. Authors are grateful to T. Glushkova for measurements of the LiF samples optical parameters.

*Corresponding author.
chekalin@isan.troitsk.ru
†Corresponding author.
dormidonov@gmail.com